\documentclass{article}

\usepackage{microtype}
\usepackage{graphicx}
\usepackage{subcaption}
\usepackage{booktabs} 
\usepackage{multirow}

\usepackage{hyperref}

\usepackage{xspace}
\usepackage[ruled,vlined,scleft]{algorithm2e}

\SetCommentSty{mycommfont}
\usepackage[preprint]{icml2026}

\usepackage{amsmath}
\usepackage{amssymb}
\usepackage{mathtools}
\usepackage{amsthm}

\theoremstyle{plain}

\theoremstyle{definition}

\theoremstyle{remark}

\newcommand{\ie}{i.e.,\ }

\newcommand{\name}{\emph{PackInfer}\xspace}

\usepackage[inline,shortlabels]{enumitem}
\newenvironment{denseitemize}{
\begin{itemize}[topsep=2pt, partopsep=0pt, leftmargin=1.5em]
  \setlength{\itemsep}{2pt}
  \setlength{\parskip}{0pt}
  \setlength{\parsep}{0pt}
}{\end{itemize}}

\hypersetup{
  colorlinks=true,      
  linkcolor=blue,       
}

\icmltitlerunning{Preprint}

\begin{document}

\twocolumn[
	\icmltitle{\name: Compute- and I/O-Efficient Attention for Batched LLM Inference}



	\icmlsetsymbol{equal}{*}
    \icmlsetsymbol{corresp}{$\dagger$}

	\begin{icmlauthorlist}
		\icmlauthor{Rui Ning}{nju,equal}
		\icmlauthor{Wei Zhang}{uiuc,equal}
		\icmlauthor{Fan Lai}{uiuc,corresp}
	\end{icmlauthorlist}

	\icmlaffiliation{nju}{Nanjing University, Nanjing, China}
    \icmlaffiliation{uiuc}{Siebel Center for Computer Science, University of Illinois Urbana-Champaign, Urbana, IL, USA}

	\icmlcorrespondingauthor{Fan Lai}{fanlai@illinois.edu}


	\vskip 0.3in
]



\printAffiliationsAndNotice{\icmlEqualContribution. \textsuperscript{$\dagger$}Corresponding author.}

\begin{abstract}
Attention efficiency is critical to large language model (LLM) inference. While prior advances optimize attention execution for individual requests (e.g., FlashAttention), production LLM serving relies on batching requests with highly heterogeneous sequence lengths for high serving throughput. 
This mismatch induces severe computation and I/O imbalance, exacerbates stragglers, and underutilizes GPU resources. 
We present \name, a kernel-level attention framework that enables compute- and I/O-aware execution for heterogeneous batched inference. \name orchestrates batched requests into load-balanced execution groups, effectively saturating GPU utilization by packing multiple requests into unified kernel launches. By constructing attention kernels directly over packed query-key regions, \name eliminates redundant computation and balances thread-block execution. It then incorporates I/O-aware grouping that co-locates shared-prefix requests and reorganizes KV caches into group-contiguous layouts, reducing memory fragmentation and redundant data movement as generation evolves.
Evaluations on real-world workloads show that \name reduces inference latency by 13.0-20.1\%, and improves throughput by 20\% compared to the state-of-the-art FlashAttention.
\end{abstract}

\section{Introduction}

Transformer-based models form the foundation of today’s large language models (LLMs), with the attention layer serving as the backbone for modeling cross-token dependencies~\cite{AttentionIsAllYouNeed, GPT3, DeepSeek}. Improving the computational and memory efficiency of attention has therefore been the key. FlashAttention~\cite{FlashAttention} reduces memory footprint by fusing kernels and recomputing intermediate results on the fly. Subsequent work further improves efficiency by exploiting structural properties of attention, including sparsity~\cite{NativeSparseAttention, DuoAttention} and KV cache sharing mechanisms~\cite{MHA, GQA}. These techniques significantly improve the efficiency of attention execution for individual requests.

However, practical serving systems routinely batch requests to maximize throughput under latency constraints~\cite{PagedAttention, SGLang}. In such settings, requests exhibit highly heterogeneous sequence lengths: many interactive queries contain only tens of tokens, while long-context requests coexist within the same batch. To enable batch execution, existing systems pad sequences to block sizes, forming structured inputs for attention kernels~\cite{PagedAttention, FlashAttention}. The heterogeneous input lengths naturally lead to imbalanced workloads across thread blocks, wasted computation cycles, and straggler effects that slow down both prefilling and decoding steps.

Existing advances, such as Flash Decoding~\cite{FlashDecoding} and Chunked-prefill~\cite{ChunkPrefill}, mitigate stragglers by splitting long requests into fixed-size chunks, enabling schedulers to interleave more uniform units of work. Despite these improvements, they remain fundamentally limited in three respects. First, execution often operates on padded blocks, wasting both computation and I/O bandwidth. Second, sequence partitioning is driven primarily by computation granularity~\cite{Prepacking}, without jointly considering memory access efficiency (e.g., for prefix or KV sharing), which limits overall performance gains. Third, chunking is applied at the granularity of individual requests and fails to exploit batch-level optimization opportunities, particularly under heterogeneous input lengths.

This paper introduces \name, a kernel-level optimization that addresses imbalanced computation and memory access arising from batching requests with heterogeneous input lengths. By rebalancing attention execution across both compute and memory dimensions, \name improves GPU utilization and reduces both time-to-first-token (TTFT) and time-between-tokens (TBT) inference latency. \name integrates with existing LLM serving stacks, such as vLLM~\cite{PagedAttention} and FlashAttention~\cite{FlashAttention}, requiring only a few lines of API-level changes, and generalizes across a wide range of transformer models.

\name addresses the aforementioned limitations through a computation- and I/O-aware packing design. It packs heterogeneous requests into length-balanced groups that enable balanced kernel computation: short requests are grouped together to avoid wasted computation, while long requests are partitioned across multiple groups, with their outputs later merged in a lossless manner consistent with FlashAttention semantics. \name further adaptively determines the number of groups to maximize end-to-end efficiency under varying workload characteristics. This packing strategy balances GPU thread-block execution, mitigating stragglers and improving overall utilization.

Beyond computation, \name jointly orchestrates the physical layout of the KV cache to ensure contiguous memory access, effectively eliminating memory fragmentation. In the presence of KV cache dependencies (e.g., prefix sharing), \name incorporates memory locality into the grouping decisions, thereby preserving data reuse while maintaining balanced execution.

In summary, this paper makes the following contributions:

\begin{denseitemize}
	\item We identify execution imbalance as a fundamental inefficiency in batched LLM inference with heterogeneous input lengths, explaining the prevalence of stragglers and poor GPU utilization in practical serving  (\S\ref{subsec:motivations}).

	\item We propose \name, a novel kernel-level inference framework built on computation- and I/O-aware token packing. \name transforms irregular batches into dense and balanced kernel executions, enabling lossless attention without altering model architectures (\S\ref{sec:design}).

	\item We evaluate \name on a wide range of real-world inference workloads and models. The results show that \name improves system throughput by 13.0--20.1\% and reduces end-to-end latency by 20\% compared to the state-of-the-art FlashAttention (\S\ref{sec:eval}).
\end{denseitemize}

\section{Background and Motivation}

\subsection{LLM Inference Execution}
\label{subsec:llm-inference}

LLM inference consists of two distinct phases: \emph{prefilling} and \emph{decoding}. Prefilling processes the entire input prompt to generate the initial KV cache, determining the Time-to-First-Token (TTFT). Since it processes all prompt tokens in parallel, prefilling is often compute-intensive~\cite{Orca, DeepSpeed}. In contrast, decoding generates tokens autoregressively: each step produces a single token while reading the entire accumulated KV cache, dictating the time-between-token (TBT). As a result, decoding exhibits low arithmetic intensity and is inherently memory-bound, making it sensitive to memory bandwidth and access efficiency~\cite{DistServe, PagedAttention}.

At the hardware level, GPU kernels execute attention by partitioning the workload into thread blocks (CTAs) that perform tiled matrix--vector or matrix--matrix operations. Each CTA is scheduled on a Streaming Multiprocessor (SM) and orchestrates data movement between high-bandwidth memory (HBM) and fast on-chip SRAM (shared memory). To maximize hardware utilization, state-of-the-art attention kernels, most notably the FlashAttention family~\cite{FlashAttention, FlashAttention2}, employ tiling and selective recomputation to keep intermediate results in SRAM, thereby avoiding the $O(n^2)$ HBM access bottleneck of naive attention.

\subsection{Motivations}
\label{subsec:motivations}

\begin{figure}[t]
	\centering
	\includegraphics[width=.9\linewidth]{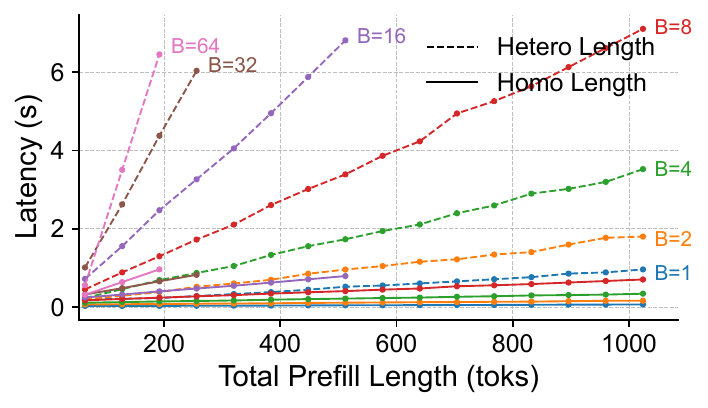}
	\vspace{-5pt}
	\caption{Heterogeneity reduces GPU utilization.}
	\label{fig:utilization} 
    \vspace{-10pt}
\end{figure}

\begin{table}[t]
    \centering
    \begin{tabular}{lrr}
        \toprule
        Metric &  Min & Max \\
        \midrule
        SM Active Cycles    & 39,886  & 1,433,315 \\
        SMSP Active Cycles  & 37,662  & 1,457,540 \\
        L1 Active Cycles    & 39,886  & 1,433,315 \\
        \bottomrule
    \end{tabular}
    \vspace{5pt}
    \caption{Context length imbalance introduces severe tail effects.}
    \label{tbl:straggler}
    \vspace{-20pt}
\end{table}

\begin{figure*}[t]
	\centering
	\includegraphics[width=0.85\linewidth]{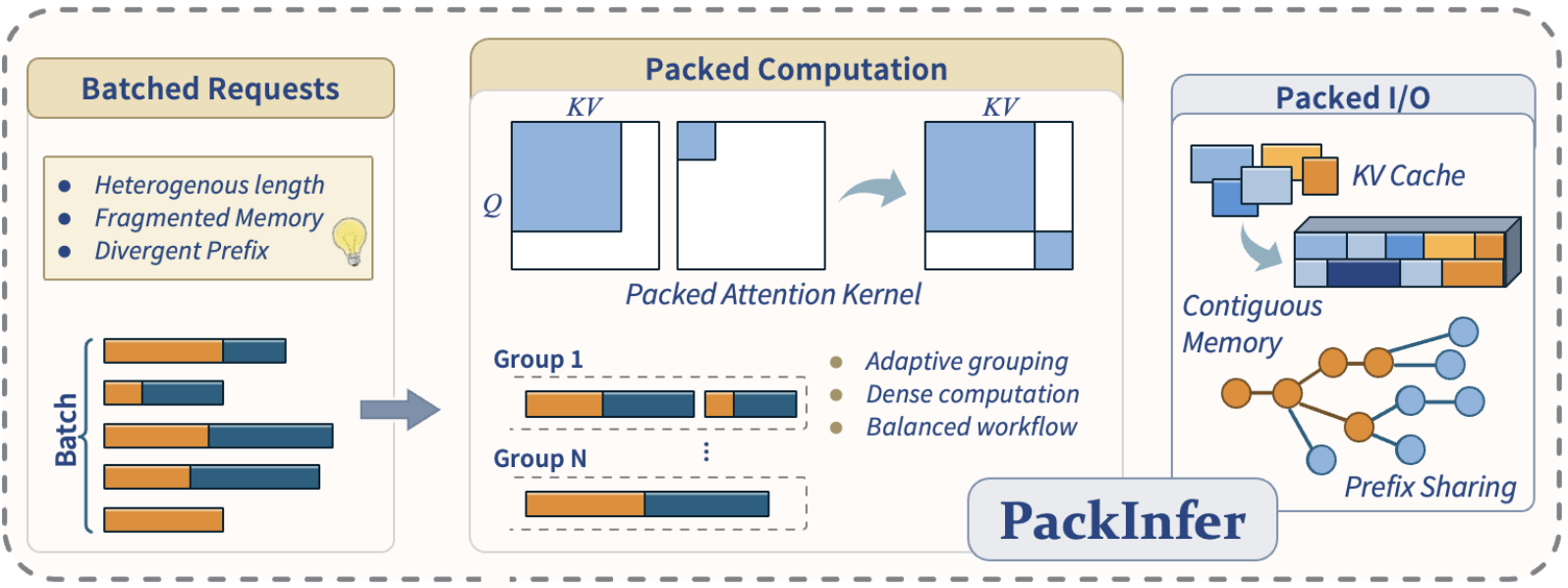}
	\caption{\name overview. It adaptively packs requests of heterogeneous input lengths into computation and I/O load-balanced groups. }
	\label{fig:overview}
	\vspace{-3pt}
\end{figure*}

\paragraph{Batch Execution under Heterogeneity Amplifies Latency.}
Batching is the dominant execution mode in production LLM serving to maximize throughput, yet batched requests often exhibit heterogeneous input lengths. Figure~\ref{fig:utilization} reports batch execution results on the Alpaca dataset using the Llama3.1-8B~\cite{llama3} model. We compare realistic heterogeneous batches against a hypothetical homogeneous baseline, where all requests have identical lengths, while the total number of input tokens per batch is the same as in the heterogeneous setting. Under heterogeneous batching, we observe a significant increase in prefilling latency; similar trends are also observed for per-step decoding latency.
First, fixed-size tiling (e.g., $128 \times 128$) forces shorter requests to execute full matrix multiplications even when most tile elements are masked. 
Second, the rigid mapping between CTAs and tiles causes resource stranding. Since each tile, even those containing a single token, occupies an entire CTA, short requests overclaim hardware resources (e.g., register memory). 

\paragraph{Workload Imbalance and Stragglers.} 
Beyond inefficiencies within individual requests, heterogeneity introduces severe temporal imbalance across GPU SMs. Because CTA allocation scales with sequence length, longer requests generate disproportionately more tiles and dominate kernel execution time. As shown in Table~\ref{tbl:straggler}, this imbalance leads to a pronounced \emph{straggler effect}: SMs assigned to short-context requests complete early and remain idle, while those processing long-context requests form the critical path. As hardware-level thread-block scheduling is tied to request-aligned tile grids, existing advances lack cross-request flexibility prevents dynamic load redistribution at the kernel level, leaving idle SM resources unable to assist straggling workloads and amplifying tail latency.

\section{\name Design}
\label{sec:design}

To enable efficient batch execution under heterogeneous LLM inference workloads, we introduce \name, a kernel-level packing framework that selectively groups requests into balanced execution units while preserving lossless attention semantics. \name explicitly accounts for I/O locality and prefix sharing, reducing memory footprint and kernel launch overhead, and applies to both the prefilling and decoding phases of transformer inference. For clarity, the remainder of this paper focuses on the attention layer, which involves both computational and memory inefficiencies.

\paragraph{Overview.}
As illustrated in Figure~\ref{fig:overview}, \name comprises two tightly integrated components: \emph{packed computation} (\S\ref{subsec:packed-Computation}) and \emph{packed I/O} (\S\ref{subsec:packed-IO}). Upon request arrival, \name analyzes the effective sequence lengths and KV access patterns of concurrent requests and groups them into load-balanced execution units subject to kernel capacity and memory constraints. These grouping decisions are further refined to account for shared KV regions (e.g., prefix sharing), explicitly grouping requests to minimize redundant memory traffic. 
Given the resulting groups, \name instantiates attention kernels whose execution domains are constructed directly from the union of valid query--key regions across requests, rather than from padded, per-request–aligned tiles. In parallel, \name reorganizes KV caches into group-specific, contiguous memory layouts that align with the packed execution domains, enabling memory coalescing.

Designing these components jointly is non-trivial. First, heterogeneous request lengths induce highly irregular execution loads, complicating the construction of compact yet correct group-level kernel execution domains. Second, inter-request dependencies, such as KV cache sharing in prefix sharing, introduce I/O coupling, making it challenging to maintain contiguous memory layouts without incurring excessive data movement or synchronization overhead. Third, these challenges evolve dynamically as generation progresses and new tokens are produced. We describe how \name addresses these challenges.

\subsection{Packed Computation}
\label{subsec:packed-Computation}

In modern attention kernels (e.g., FlashAttention), the computation space is discretized into fixed-size tiles of dimension $T \times T$ to optimize for SRAM residency and coalesced memory access. Under conventional batching, each request $i$ with effective sequence length $L_i$ is assigned a dedicated computation region aligned to this tile size $T$, resulting in a per-request utilization $\eta_i = \frac{L_i^2}{T^2}$, 
where the $L_i^2$ term reflects the quadratic complexity of self-attention. In practice, the tile size $T$ is fixed to hardware-aligned thresholds (typically 128 or 256) to ensure efficient memory coalescing and high utilization of GPU tensor cores.

\begin{figure}
    \centering
    \includegraphics[width=.7\linewidth]{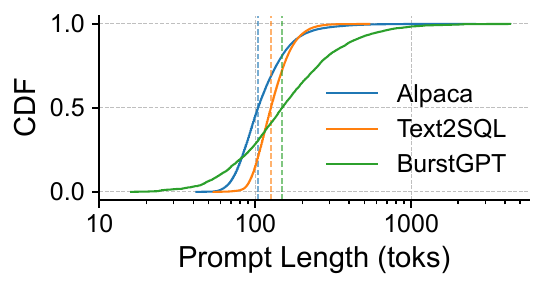}
    \caption{Real-world inference workloads exhibit highly heterogeneous request lengths, with many shorter than 128 tokens.
    }
    \label{fig:context_length_cdf}
    \vspace{-15pt}
\end{figure}

However, real-world inference workloads exhibit highly skewed, long-tailed sequence length distributions. As shown in Figure~\ref{fig:context_length_cdf}, more than 60\% of requests have input lengths shorter than 128 tokens. When $L_i \ll T$, a large fraction of each $T \times T$ tile is occupied by padding or masked elements, leading to wasted memory bandwidth and idle compute cycles.

To maximize hardware occupancy, \name partitions requests in a batch $\mathcal{R}$ into $G$ disjoint packed groups $\{\mathcal{S}_1, \dots, \mathcal{S}_G\}$, where each group is executed by a single kernel invocation. For a packed group $\mathcal{S}_g$, the collective utilization $\eta(\mathcal{S}_g)$ is defined as the ratio between the effective computation contributed by all requests in the group and the total tiled capacity:
\begin{equation}
\eta(\mathcal{S}_g) = \frac{\sum_{i \in \mathcal{S}_g} L_i^2}{T^2}
\quad \Rightarrow \quad
\eta_{\text{batch}} = \frac{\sum_{i=1}^{N} L_i^2}{G \cdot T^2}.
\end{equation}
Compared to naive execution, where each request is assigned its own kernel invocation ($G = N$), packing reduces the total tiled capacity $G \cdot T^2$ by co-locating multiple requests within a single kernel of size $T$. Notably, while the aggregate utilization in the formula above is determined by the number of groups $G$, it is invariant to the specific distribution of requests within those groups. To further unlock hardware potential, we must look beyond intra-kernel packing and further address the heterogeneity.

\paragraph{Inter-group Workload Balancing}
While packing maximizes intra-kernel efficiency, GPU execution can suffer from bubbles if the workloads $L(\mathcal{S}_g) = \sum_{i \in \mathcal{S}_g} L_i$ are highly heterogeneous across groups. Specifically, SMs assigned to groups with smaller aggregate lengths will finish their execution early, but remain idle until the longest group wave reaches its synchronization barrier. 

To mitigate such imbalance, we first define a group $\mathcal{S}_g$ as feasible if it satisfies the kernel's capacity and system memory constraints:
\begin{equation}
	\Phi(\mathcal{S}_g) \triangleq \left( \sum_{i \in \mathcal{S}_g} L_i \le \mathcal{C} \right) \wedge \left( M(\mathcal{S}_g) \le M_{\max} \right)
\end{equation}
where $\mathcal{C}$ denotes the maximum total token length per group, decided by profiling and latency requirements, and $M(\mathcal{S}_g)$ accounts for the KV cache and required workspace buffers. Under these constraints, the grouping objective can be formulated as minimizing the discrepancy in total lengths across groups:
\begin{equation}
\small
	\min_{\{\mathcal{S}_g\}} \; | \max_{g} L(\mathcal{S}_g) - \min_{g} L(\mathcal{S}_g) | \quad \text{s.t.} \quad \Phi(\mathcal{S}_g) = \text{True}, \forall g
\end{equation}
The partitioning problem is a variant of the NP-hard bin-packing problem~\cite{JITServe}, where finding a global optimum in real-time decoding is computationally prohibitive, especially for online serving deployments.

\begin{algorithm}[t]
	\caption{Packed Computation and I/O Scheduling} 
	\label{alg:packed-grouping}
	\LinesNumbered
	\SetNlSty{textbf}{}{}
	\SetNlSkip{-0.5em}
	\setlength{\algomargin}{2.5em}
	\SetInd{1.5em}{1.5em}
	\KwIn{Set of requests $\mathcal{R}$, Group capacity $\mathcal{C}$ from profiling, Paged KV Cache $\mathcal{M}_{paged}$}
	\KwOut{Grouped partitions $\{\mathcal{S}_g\}$, Contiguous buffers $\{\mathcal{B}_g\}$, and Offset tables $\{\mathcal{O}_g\}$.}

	\vspace{3pt}

	\tcp{Part 1: Inter-group Workload Balancing}

	\quad $L_{total} \leftarrow \sum_{i \in \mathcal{R}} L_i, \ G \leftarrow \lceil L_{total} / \mathcal{C} \rceil$

	\label{algo:estimate_group_size}
	\quad Initialize $G$ groups $\{\mathcal{S}_1, \dots, \mathcal{S}_G\}$ and their cumulative lengths $L(\mathcal{S}_g) \leftarrow 0$\;

	\quad Sort $\mathcal{R}$ in descending order of effective length $L_i$\;
	\label{algo:sort_requests}

	\quad \ForEach{request $i \in \mathcal{R}$}{
		\tcp{Find the least loaded group}
		$g^* \leftarrow \arg\min_g L(\mathcal{S}_g)$ \\
		\label{algo:group-assgin-start}
		\If{$\Phi(\mathcal{S}_{g^*} \cup \{i\}) = \text{True}$}{
			$\mathcal{S}_{g^*} \leftarrow \mathcal{S}_{g^*} \cup \{i\}, \
				\label{algo:group-assgin-end}
				L(\mathcal{S}_{g^*}) \leftarrow L(\mathcal{S}_{g^*}) + L_i$\;
		}
		\Else{\label{algo:else-start}
			$\mathcal{S}_{G+1} \leftarrow \{i\}, \ G \leftarrow G + 1$\;
		}
		\label{algo:else-end}
	}

    \vspace{3pt}
    \tcp{Part 2: Packed I/O Layout Preparation}
    \quad \ForEach{group $\mathcal{S}_g \in \{\mathcal{S}_1, \dots, \mathcal{S}_G\}$}{
        \tcp{Identify shared prefixes $\mathcal{P}$ and associated suffixes $\mathcal{Q}$}
        $\{(\mathcal{P}_k, \{\mathcal{Q}_{i,k}\})\} \leftarrow \text{TriePartition}(\mathcal{S}_g), \  \Delta \leftarrow 0$\;
        \label{algo:traverse}
        
        \ForEach{prefix $\mathcal{P}_k$ and its suffix set $\{\mathcal{Q}_{i,k}\}$}{
            $\text{Copy}(\mathcal{M}_{paged}[\mathcal{P}_k] \to \mathcal{B}_g)$
            \label{algo:copy-start}
            \BlankLine
            ${\Delta}_{prefix} \leftarrow \Delta, \ \Delta \leftarrow \Delta + L_{\mathcal{P}_k}$\;
            
            \ForEach{suffix $\mathcal{Q}_{i} \in \{\mathcal{Q}_{i,k}\}$}{
                $\text{Copy}(\mathcal{M}_{paged}[\mathcal{Q}_{i}] \to \mathcal{B}_g)$  \BlankLine 
                $\mathcal{O}_g[i] \leftarrow (\Delta_{prefix}, L_{\mathcal{P}_k}, \Delta, L_{\mathcal{Q}_{i}})$\; \BlankLine 
                $\Delta \leftarrow \Delta + L_{\mathcal{Q}_{i}}$\;
            }
            \label{algo:copy-end}
        }
    }
    \quad \Return $\{\mathcal{S}_g, \mathcal{B}_g, \mathcal{O}_g\}$\;
\end{algorithm}

To solve this problem efficiently at runtime, we employ a lightweight greedy grouping algorithm (Algorithm~\ref{alg:packed-grouping}). Given the incoming requests and the group capacity $\mathcal{C}$, \name first estimates the target number of groups based on the aggregate request length to encourage balanced workloads across groups (Line~\ref{algo:estimate_group_size}). The requests are then sorted in descending order of effective length $L_i$ (Line~\ref{algo:sort_requests}). We iteratively assign each request to the group $\mathcal{S}_{g^*}$ with the current minimum cumulative length $L(\mathcal{S}_{g^*})$, provided that the feasibility constraint $\Phi(\mathcal{S}_{g^*} \cup \{i\})$ is satisfied (Lines~\ref{algo:group-assgin-start}-\ref{algo:group-assgin-end}). 
If the constraint is violated, a new group is initialized to accommodate the request (Lines~\ref{algo:else-start}-\ref{algo:else-end}). This strategy transforms heterogeneous workloads into homogeneous execution groups, synchronizing the completion times of concurrent hardware streams and maximizing the global hardware duty cycle.

\paragraph{Adaptive Grouping.}
The effectiveness of the grouping algorithm depends on the group capacity $\mathcal{C}$, defined as the maximum total token length of a group that sustains high hardware efficiency for a given GPU and kernel configuration. Because the computational cost of attention depends only on sequence length rather than request content, $\mathcal{C}$ can be determined through lightweight profiling.

We first perform offline profiling over a range of group sequence lengths to identify capacity thresholds that achieve near-peak efficiency. To adapt to workload shifts (e.g., changes in input length distributions), \name further refines $\mathcal{C}$ using online profiling driven by lightweight runtime signals such as observed latency and throughput. This online profiling incurs negligible overhead: each decoding step naturally yields one performance sample, allowing the impact of different group capacities to be explored within a few dozen decoding steps and amortized across millions of daily inference requests.

Although the initial greedy assignment minimizes inter-group workload variance, residual imbalance can accumulate as requests generate new tokens during decoding. To prevent workload divergence, \name selectively triggers regrouping based on observed drift. We define the per-step drift as
$
\Delta L = \left| \max_g L(\mathcal{S}_g) - \min_g L(\mathcal{S}_g) \right|,
$
and trigger regrouping when the cumulative imbalance over $t$ decoding steps exceeds a fraction of the group capacity:
\begin{equation}
t \cdot \Delta L \ge \frac{\mathcal{C}}{2}
\end{equation}
The threshold $\mathcal{C}/2$ balances regrouping overhead against recovered utilization. At this point, the aggregate execution bubbles induced by inter-group imbalance approach half of a full execution unit, equivalent to the bubble degree of creating a new group, and thus making regrouping more beneficial as we will continue generating more tokens. In practice, since the group capacity $\mathcal{C}$ (e.g., 8192 tokens) is typically much larger than the per-step token growth of individual requests, this condition is reached only every 20--40 decoding steps, maintaining high hardware efficiency with minimal regrouping overhead~(\S\ref{eval:e2e}).

\subsection{Packed IO}
\label{subsec:packed-IO}

While packed computation enhances compute utilization, the memory often becomes another bottleneck due to fragmented access and redundant data movement, especially for the decoding stage or prefix sharing. To address this, we design a packed I/O strategy, which streamlines memory layouts and leverages data locality.

\begin{figure}
    \centering
    \includegraphics[width=\linewidth]{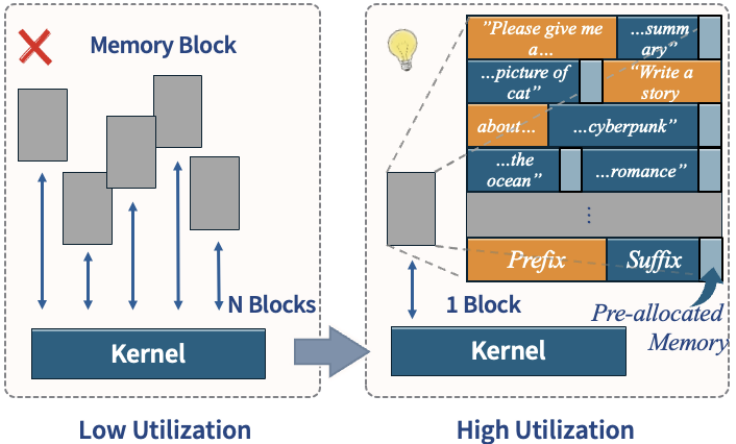}
    \caption{Contiguous Memory Consolidation in \name. \name re-aligns scattered KV cache blocks into a unified, high-density buffer to maximize I/O throughput. A pre-allocated headroom is reserved for each request to accommodate future tokens.}
    \label{fig:pack-io}
\end{figure}

\paragraph{Prefix-Aware Grouping and Reuse.}
In practical workloads, many requests share common prompt prefixes, such as system prompts or conversational histories. While existing advances~\cite{SGLang} enable prefix sharing via physical memory mapping, they treat requests in a batch independently at the kernel level. This means the prefix KV cache must be fetched into registers or Shared Memory multiple times, leading to avoidable memory contention and cache thrashing when multiple warps compete for the same cache lines simultaneously.

We observe that prefix sharing can be naturally integrated into the \name framework by representing the requests within a group $\mathcal{S}_g$ as a shared prefix tree $\mathcal{T}_g$. By traversing $\mathcal{T}_g$, we identify the set of unique common prefixes $\{\mathcal{P}_k\}$ and their corresponding suffix sets $\{\mathcal{Q}_{i,k}\}$ (Line~\ref{algo:traverse}). We then reorganize the physical KV cache into a prefix-first layout, where shared segments are stored once and followed by their respective suffixes (Fig~\ref{fig:pack-io}). Formally, the total I/O volume $M(\mathcal{S}_g)$ for fetching the KV cache is reduced from the naive sum $\sum_{i \in \mathcal{S}_g} (L_{\mathcal{P}_i} + L_{\mathcal{Q}_i})$ to:
\begin{equation}
M(\mathcal{S}_g) = \sum_{\mathcal{P}_k \in \text{Prefixes}(\mathcal{T}_g)} \left( L_{\mathcal{P}_k} + \sum_{\mathcal{Q}_{i,k} \in \text{Suffixes}(\mathcal{P}_k)} L_{\mathcal{Q}_{i,k}} \right)
\end{equation}
Importantly, integrating prefix sharing at the kernel level maintains the integrity of our inter-group balancing strategy. During the partitioning phase, we redefines a request's effective contribution to a group based on its unique suffix length relative to that group's existing prefixes, \ie, $\hat{L}_i = L_i - L_{shared, i}^g$ if the request is assigned to group $g$, which our greedy grouping algorithm will automatically support in packing (Line~\ref{algo:sort_requests}-Line~\ref{algo:else-end}). Additionally, the total workload of a group is no longer a naive sum but is calculated by deducting the redundant prefix lengths, ensuring the group capacity $\mathcal{C}$ is utilized only by unique token data.

\paragraph{Contiguous Memory Consolidation}
Recent advances such as PagedAttention~\cite{PagedAttention} optimize KV cache management by partitioning memory into non-contiguous blocks, effectively reducing external fragmentation and supporting dynamic sequence growth. While effective for standalone requests, this design introduces significant overhead under packed execution. When a single kernel concurrently indexes KV blocks across multiple heterogeneous requests, the fragmented layout induces irregular memory access patterns, leading to poor bandwidth utilization and increased SM overhead due to divergent address translation and pointer chasing across request groups.

To address this inefficiency, \name introduces a contiguous memory consolidation phase prior to kernel invocation. This phase gathers active KV cache entries from disjoint pages and compacts them into a unified, high-density memory buffer aligned with the packed execution domain. Beyond simple data movement, the consolidation process eliminates internal fragmentation by copying only valid token states into the workspace (Lines~\ref{algo:copy-start}--\ref{algo:copy-end}). To amortize consolidation cost across decoding steps, we employ a proactive allocation strategy with a suffix headroom parameter~$\delta$. For each request $\mathcal{Q}_i$, the allocated memory size is
$
M(\mathcal{Q}_i) = L_{\mathcal{Q}_i} + \delta
$. 
which reserves space for future token generation and allows multiple decoding iterations to proceed without triggering re-alignment. By transforming sparse, padding-heavy accesses into dense, contiguous memory streams, \name improves memory coalescing, maximizes effective bandwidth utilization, and significantly reduces I/O pressure during packed execution.

\section{Experiment}
\label{sec:eval}

\subsection{Experiment setup}
We implement \name as a modular library consisting of approximately 1,000 lines of highly optimized CUDA and C++ kernels. The core is encapsulated within a drop-in replacement for the standard FlashAttention API, ensuring broad applicability and ease of integration.  We show this versatility by integrating \name with Nano-vLLM~\cite{nanovllm2024}, a streamlined version of the vLLM inference engine supporting diverse existing serving frameworks.


\paragraph{Testbed and Workloads.}
We conduct experiments on a single NVIDIA A100 GPU with 40\,GB memory by default, and extend to 4$\times$A100 GPUs for distributed evaluations. We further validate the robustness of \name across different hardware platforms, including NVIDIA H200 and A40 GPUs. We evaluate \name using three representative LLMs, Qwen3-4B~\cite{Qwen3technicalreport}, Mistral-7B~\cite{Mistral7b}, and Qwen3-30B-A3B (MoE) variants, under three real-world request traces: Alpaca (instruction-following), LMSYS (realistic ChatGPT workloads), and Text2SQL (query generation). Each workload contains between 7{,}000 and 200{,}000 prompts. Requests are scheduled using the default vLLM scheduling policy, first-come-first-served (FCFS), across all settings.

\paragraph{Baselines.}
We compare \name against state-of-the-art attention execution approaches, including FlashAttention~\cite{FlashAttention2}, and Prepack~\cite{Prepacking}, a recent KV-cache reordering advance that improves memory layout without modifying the attention kernel.

\paragraph{Metrics.}
We adopt standard LLM serving metrics that capture both user-facing latency and hardware efficiency. Latency metrics include time-to-first-token (TTFT), time-between-token (TBT), and total time-to-last-token (TTLT). We also report GPU utilization metrics such as SM occupancy and memory bandwidth. All metrics are computed over the full trace using identical inference parameters. 

Results are averaged over five independent runs.

\subsection{End-to-End Performance}
\label{eval:e2e}

\begin{figure}[t]
	\centering
	\begin{subfigure}[t]{0.49\textwidth}
		\centering
		\includegraphics[width=\linewidth]{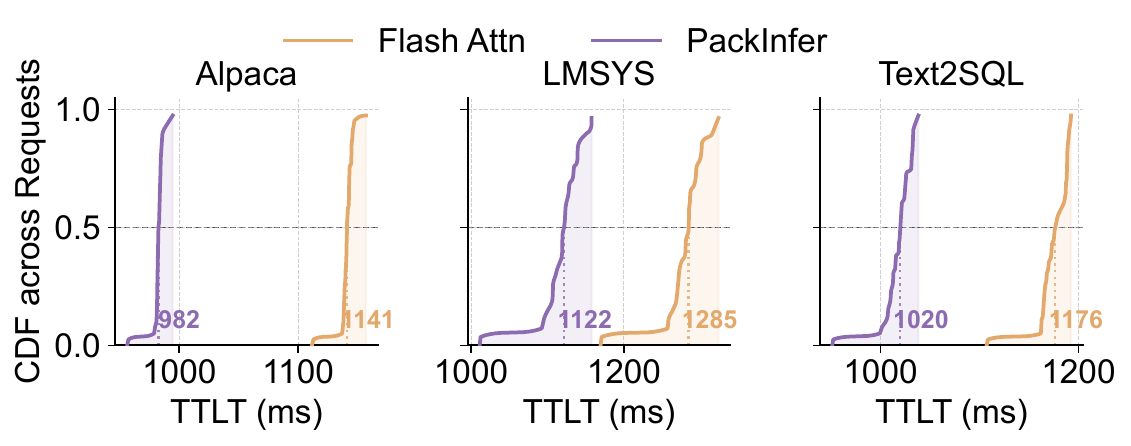}
		\caption{Mistral-7B TTLT latency.}
	\end{subfigure}
	\hfill
	\begin{subfigure}[t]{0.49\textwidth}
		\centering
		\includegraphics[width=\linewidth]{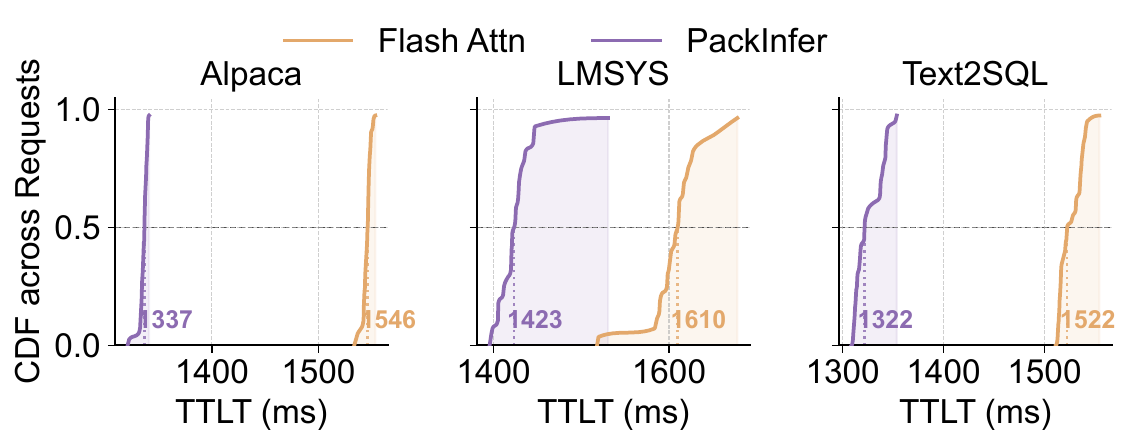}
		\caption{Qwen3-4B TTLT latency.}
	\end{subfigure}
	\caption{\name improves time-to-last-token (TTLT) latency.}
	\label{fig:eval-ttlt}
\end{figure}

\begin{figure}[t]
	\centering
	\begin{subfigure}[t]{0.49\textwidth}
		\centering
		\includegraphics[width=\linewidth]{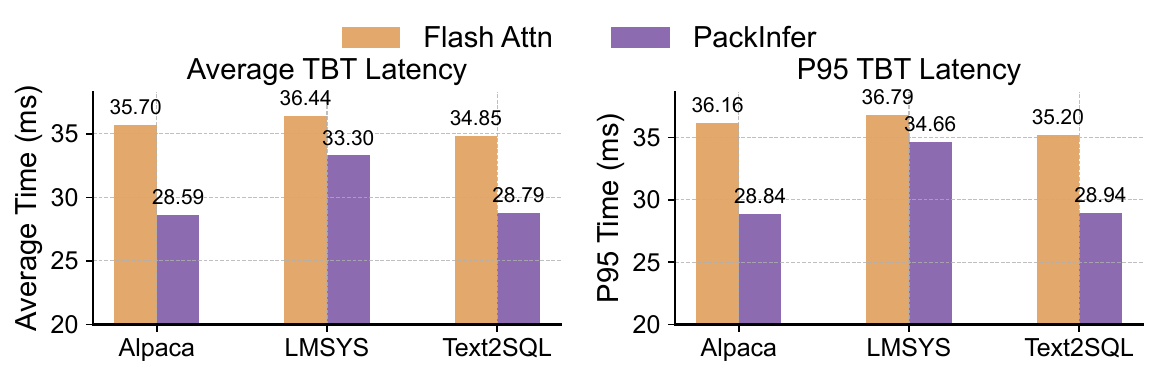}
		\caption{Mistral-7B TBT latency.}
	\end{subfigure}
	\hfill
	\begin{subfigure}[t]{0.49\textwidth}
		\centering
		\includegraphics[width=\linewidth]{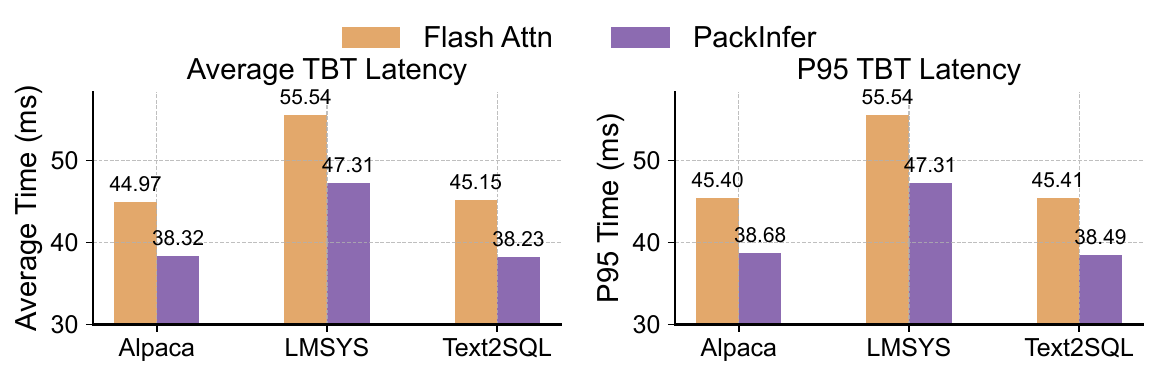}
		\caption{Qwen3-4B TBT latency.}
	\end{subfigure}
	\caption{\name improves time-between-tokens (TBT) latency.}
	\label{fig:eval-tbt}
\end{figure}

\begin{figure}[t]
	\begin{subfigure}[t]{0.235\textwidth}
		\centering
		\includegraphics[width=\linewidth]{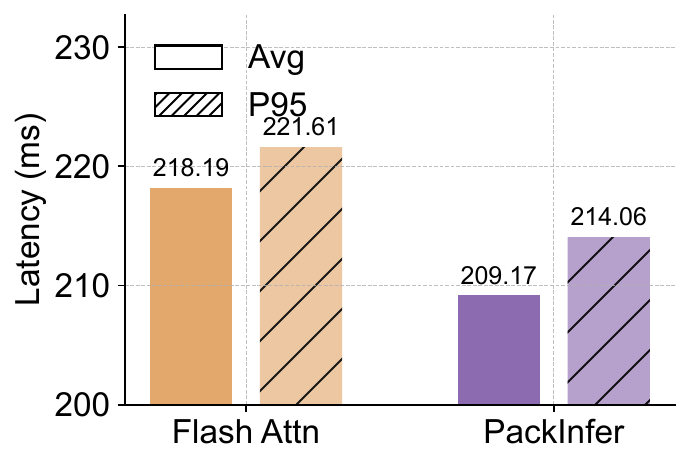}
		\caption{Qwen3-30B-A3B TBT.}
	\end{subfigure}
	\begin{subfigure}[t]{0.235\textwidth}
		\centering
		\includegraphics[width=\linewidth]{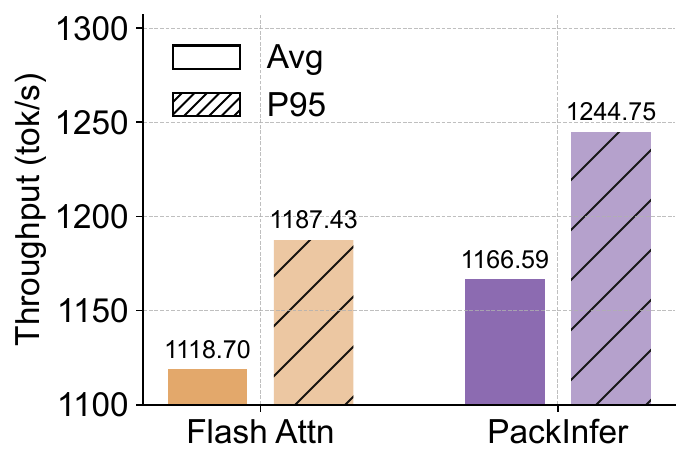}
		\caption{Qwen3-30B-A3B Throughput.}
	\end{subfigure}
    \caption{\name improves serving performance for MoE model.}
    \label{fig:eval-latency-moe}
\end{figure}

\paragraph{\name improves serving latency.}
Figure~\ref{fig:eval-ttlt} and Figure~\ref{fig:eval-tbt} show that \name achieves substantial latency reductions, improving TBT by 13.0--20.1\% and total TTLT by 3.8--20.2\%. Table~\ref{table:ttft} further shows that \name reduces TTFT latency by up to 18.6\% compared to the state-of-the-art FlashAttention baseline, without introducing additional overhead in existing deployments. These gains primarily stem from mitigating I/O bottlenecks and packing attention computation to reduce kernel-level stragglers, thereby improving execution balance across GPU SMs. As shown in Figure~\ref{fig:eval-latency-moe}, this efficiency gain persists on MoE-based models, confirming \name's generalized speedups across model architectures and sizes.

\begin{table}[t]
	\centering
	\small
		\begin{tabular}{l l rr rr}
		\toprule
		 &  & \multicolumn{2}{c}{Qwen3-4B} & \multicolumn{2}{c}{Mistral-7B} \\
		\cmidrule(lr){3-4} \cmidrule(lr){5-6}
		Dataset & Backend & Avg & P99 & Avg & P99 \\
		\midrule
		\multirow{3}{*}{Alpaca} 
        & Prepack    & 96.7 &1158.3 & 94.6 &1118.7 \\
		& Flash Attn & 75.3 & 506.4 & 93.8 & 535.6 \\
		& \name      & \textbf{61.4} & \textbf{497.3} & \textbf{79.9} & \textbf{482.1} \\
        \midrule
		\multirow{3}{*}{LMSYS}
        & Prepack    & 426.0 &4834.0 & 397.3 &3836.3 \\
		& Flash Attn & 252.9 & 761.2 & 327.8 & 823.6 \\
		& \name      & \textbf{247.1} & \textbf{750.8} & \textbf{324.5} & \textbf{815.2} \\
        \midrule
		\multirow{3}{*}{Text2SQL} 
        & Prepack    & 125.3 &1173.3 &157.7 & 1175.0 \\
		& Flash Attn & 108.9 & 520.2 & 159.9 & 603.4 \\
		& \name      & \textbf{97.8} & \textbf{514.5} & \textbf{150.8} & \textbf{563.0} \\
		\bottomrule
	\end{tabular}
    \vspace{2pt}
    \caption{\name achieves better time-to-first-token (TTFT) latency (ms) across datasets and backends.}
    \label{table:ttft}
    \vspace{-10pt}
\end{table}

\begin{figure}[t]
	\centering
	\begin{subfigure}[t]{0.235\textwidth}
		\centering
		\includegraphics[width=\linewidth]{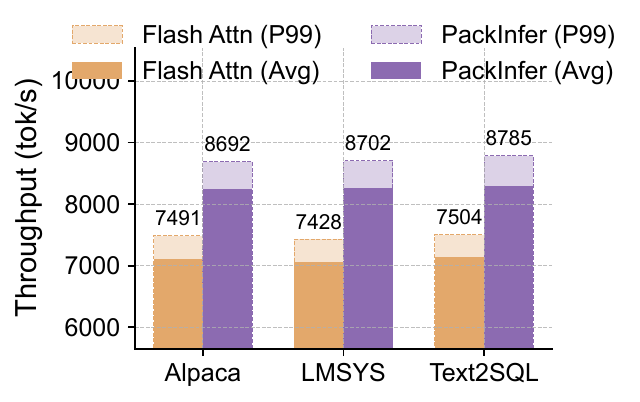}
		\caption{Mistral Throughput.}
	\end{subfigure}
	\begin{subfigure}[t]{0.235\textwidth}
		\centering
		\includegraphics[width=\linewidth]{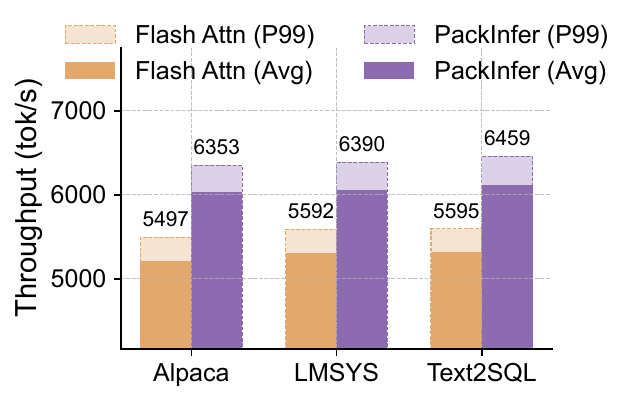}
		\caption{Qwen3 Throughput.}
	\end{subfigure}
	\caption{End-to-end throughput on Mistral and Qwen3 workloads, where P99 denotes the upper bound of throughput.}
	\label{fig:eval-tpt}
\end{figure}

\vspace{-3pt}

\begin{table}[t]
	\centering
	\small
	\caption{Kernel utilization metrics.}
	\label{tab:utilization}
	\begin{tabular}{lcc}
		\toprule
		Backend & Tensor Core Tpt. & SM Inst. Issue Tpt. \\
		\midrule
		Flash Attn & 18\% & 17\% \\
		\name & 33\% & 25\% \\
		\bottomrule
	\end{tabular}
\end{table}

\paragraph{\name improves serving throughput.}
Figure~\ref{fig:eval-tpt} reports that compared to FlashAttention, \name delivers 9.3--24.9\% higher token throughput. The advantage of \name scales with the degree of workload heterogeneity, as our design minimizes memory access overheads and maximizes computational occupancy through request packing.

\paragraph{\name improves resource utilization.}
Table~\ref{tab:utilization} reports low-level hardware utilization metrics. Consistent with our analysis (\S\ref{subsec:packed-Computation}), \name improves Tensor Core utilization and SM instruction issue throughput by 8--15\%. These results confirm that our packing-based design effectively translates reduced computational and I/O waste into higher sustained hardware efficiency.





\subsection{Ablation Studies}

We next perform ablation studies on Qwen3-4B model by default to show \name's robustness in improvements. 

\begin{figure}[t]
	\centering
	\includegraphics[width=.7\linewidth]{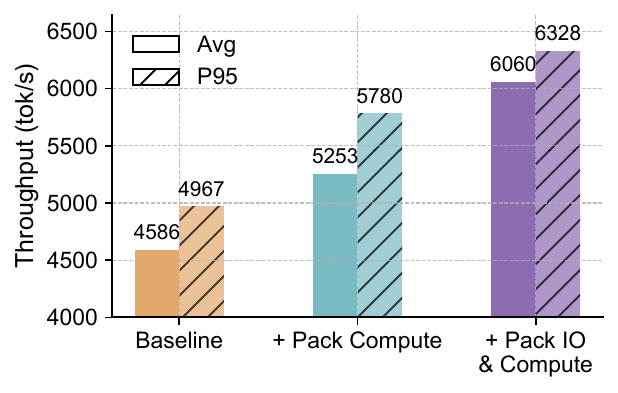}
    \vspace{-5pt}
	\caption{Performance Breakdown by components.}
	\label{fig:breakdown}
\end{figure}

\paragraph{Performance Breakdown by Components.}
To understand the sources of performance gains, we decouple the contributions of \emph{Packed Computation} and \emph{Packed I/O}. As shown in Figure~\ref{fig:breakdown}, both components independently improve performance over the FlashAttention baseline. Packed Computation naturally delivers a more pronounced speedup because it reduces kernel invocations overhead and recovers the computational cycles wasted by highly skewed sequence length distributions. Nevertheless, Packed IO provides additional gains by minimizing redundant memory traffic in end-to-end inference, yielding complementary benefits.

\begin{figure}[t]
	\centering
	\includegraphics[width=\linewidth]{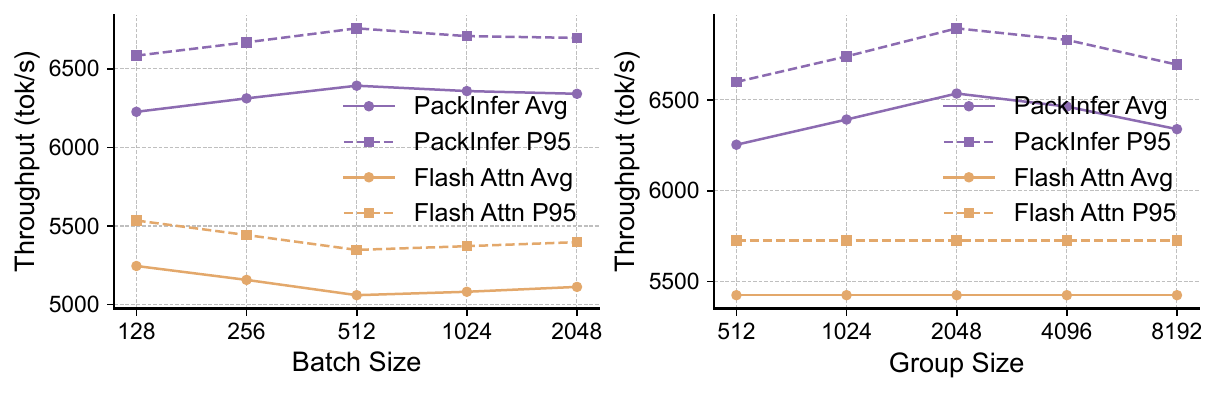}
	\caption{\name consistently outperforms across group size and batch size settings. 
    }
	\label{fig:ablation-group-batch}
\end{figure}

\paragraph{Impact of Batch Size and Group Size.}
We study the robustness of \name by sweeping both batch size and group size (Figure~\ref{fig:ablation-group-batch}). Across all batch sizes, \name consistently achieves 1.2--1.3$\times$ higher throughput than the FlashAttention baseline. Notably, the average throughput of \name even exceeds the P95 throughput of FlashAttention across  batch size settings, highlighting its effectiveness.

With respect to group size, \name exhibits a convex performance trend, reaching peak throughput at a group size of 2048. This operating point represents the optimal trade-off between improved hardware utilization from aggressive packing and increased internal fragmentation at larger group sizes, validating the effectiveness of our offline profiling–guided group capacity selection. 

\begin{figure}[t]
    \begin{minipage}{0.23\textwidth}
        \centering
        \includegraphics[width=\linewidth]{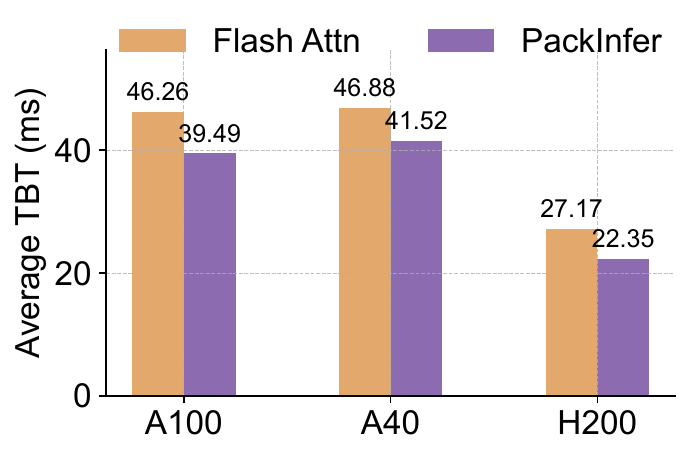}
        \caption{\name achieves improvements across hardware.}
        \label{fig:gpu-architecture}
    \end{minipage}
    \hfill
    \begin{minipage}{0.23\textwidth}
        \centering
        \includegraphics[width=\linewidth]{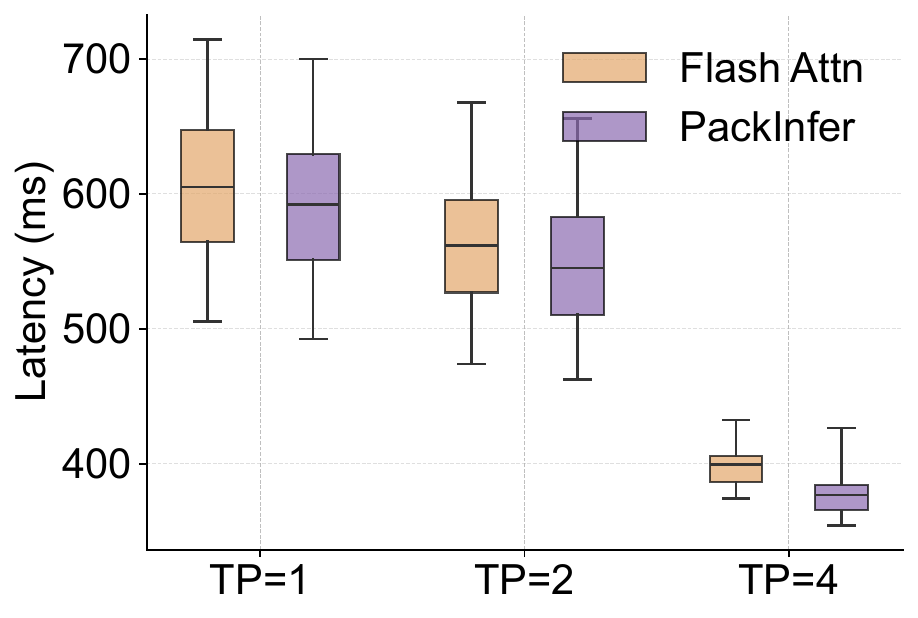}
        \caption{\name achieves gains in multi-GPU settings.}
        \label{fig:tp}
    \end{minipage}
\end{figure}

\paragraph{Impact of Hardware.}
Using the same workloads, Figure~\ref{fig:gpu-architecture} shows that \name reduces TBT latency by 11--19\% across H100, A100, and A40 GPUs. These results demonstrate that \name generalizes well across diverse GPU architectures. Moreover, \name integrates seamlessly with existing inference applications and frameworks, requiring only a few lines of API-level changes.


\paragraph{Support for Distributed Execution.} 
\name is designed to naturally support multi-GPU distributed inference. Figure~\ref{fig:tp} reports TTFT under different tensor parallelism (TP) degrees using the Qwen3-32B model. The results show that \name consistently delivers substantial performance improvements in both single-GPU (TP=1) and multi-GPU (TP=4 on 4$\times$A100) configurations. Importantly, the packing mechanism scales gracefully with increasing TP degree, demonstrating that the efficiency gains achieved on a single device carry over to distributed settings without introducing additional synchronization or communication bottlenecks.

\section{Related Work}

\paragraph{Attention Kernel Optimization.}
FlashAttention and its successors~\cite{FlashAttention, FlashAttention2} improve training and prefilling efficiency by combining block-wise tiling with selective recomputation to avoid materializing the attention matrix. For long-context inference, sparse attention techniques~\cite{AttentionSink, DuoAttention, H2O, SparseVideoGen} exploit token-level importance to reduce computation and KV access, while kernel-level quantization approaches such as SageAttention~\cite{SageAttention, SageAttention2} accelerate attention by applying low-bit arithmetic to the $QK^\top$ computation. Architectural variants including Multi-Query and Grouped-Query Attention~\cite{MHA, GQA} further reduce KV-cache memory traffic during decoding. These approaches primarily optimize individual attention operators under homogeneous execution assumptions, whereas \name addresses the complementary challenge due to heterogeneous batched inference workloads.

\paragraph{KV Cache Management.}
Beyond kernel design, existing advances have focused on system-level inference efficiency, particularly KV cache management across prefilling and decoding~\cite{PagedAttention, FlexGen}. Token-level techniques reduce memory footprint through selective retention~\cite{FastGen, H2O}, budgeted allocation~\cite{PyramidKV, AdaKV}, or quantization~\cite{FlexGen, KVQuant}. At the system level, PagedAttention~\cite{PagedAttention} eliminates external fragmentation by managing KV caches as non-contiguous blocks, while prefix-aware caching~\cite{SGLang} reduces redundant computation for shared prompts. These works primarily optimize KV storage, placement, or scheduling, whereas \name jointly reorganizes KV layouts and kernel execution domains.

\paragraph{Request Batching and Scheduling.}
Efficient request batching and scheduling are essential for meeting service-level objectives in LLM inference. Orca~\cite{Orca} introduces iteration-level scheduling (continuous batching). Recent length-aware schedulers~\cite{JITServe} mitigate this by predicting generation lengths and approximating Shortest-Job-First~\cite{LTR} or Multi-Level Feedback Queue~\cite{FastServe} policies, often combined with preemption or dynamic partitioning between prefilling and decoding. \name operates at the kernel execution level and facilitates scheduling policies.

\section{Conclusion}
\vspace{-2pt}
This paper introduces \name, a kernel-level packing framework that enables efficient attention execution for batched LLM inference under heterogeneous workloads. \name constructs padding-free attention kernels, balances GPU thread-block utilization across requests, and reorganizes KV cache layouts to improve memory locality while preserving lossless attention semantics. Evaluations on real-world workloads show that \name improves inference latency by 13.0-20.1\% and throughput by 20\% compared to the state-of-the-art FlashAttention.


\bibliography{ref}
\bibliographystyle{icml2026}

\newpage
\appendix
\onecolumn
\section{Notations}
\label{appendix:notations}
Table~\ref{tab:notations} summarizes the key notations and symbols used in the description of \name.
\begin{table}[ht]
\centering
\caption{Summary of core notations used in \name.}
\label{tab:notations}
\small
\begin{tabular}{llc}
\toprule
\textbf{Symbol} & \textbf{Definition} & \textbf{Domain / Units} \\
\midrule

\multicolumn{3}{l}{\textbf{Workload \& Hardware Parameters}} \\

$\mathcal{R}$ 
& Total set of concurrent inference requests in a batch 
& Set \\

$N$ 
& Number of requests in a batch ($N = |\mathcal{R}|$) 
& $\mathbb{Z}^+$ \\

$L_i$ 
& Effective sequence length of request $i$ 
& tokens \\

$T$ 
& Fixed tile size of the attention kernel (e.g., 128, 256) 
& tokens \\

$\mathcal{C}$ 
& Group capacity (max total token length per group) 
& tokens \\

$\delta$ 
& Suffix headroom reserved for future token generation 
& tokens \\

\midrule
\multicolumn{3}{l}{\textbf{Grouping \& Scheduling}} \\

$G$ 
& Number of disjoint packed groups 
& $\mathbb{Z}^+$ \\

$\mathcal{S}_g$ 
& The $g$-th packed group (subset of $\mathcal{R}$) 
& Set \\

$L(\mathcal{S}_g)$ 
& Aggregate sequence length of group $\mathcal{S}_g$ 
& tokens \\

$\eta(\mathcal{S}_g)$ 
& Collective hardware utilization of group $\mathcal{S}_g$ 
& $[0, 1]$ \\

$\Phi(\mathcal{S}_g)$ 
& Feasibility constraint function for a group 
& \{True, False\} \\

$\Delta L$ 
& Inter-group workload drift (imbalance) 
& tokens \\

\midrule
\multicolumn{3}{l}{\textbf{Prefix Sharing \& I/O}} \\

$\mathcal{T}_g$ 
& Prefix tree representing shared sequences in group $g$ 
& Tree structure \\

$\mathcal{P}_k$ 
& The $k$-th unique common prefix 
& Sequence \\

$\mathcal{Q}_{i,k}$ 
& Suffix of request $i$ associated with prefix $\mathcal{P}_k$ 
& Sequence \\

$M(\mathcal{S}_g)$ 
& Total memory I/O volume for fetching KV cache of group $g$ 
& Bytes / tokens \\

$\mathcal{B}_g$ 
& Contiguous memory buffer for group $g$ 
& Buffer address \\

$\mathcal{O}_g$ 
& Offset table for mapping request segments to $\mathcal{B}_g$ 
& Table \\

\bottomrule
\end{tabular}
\end{table}

\section{Detailed Experimental Settings}
\label{appendix:experimental_settings}
To ensure the reproducibility of our results, we provide comprehensive details regarding our hardware, software environment, hyperparameter configurations, and benchmark setups.

\subsection{Hardware and Software Environment}
Our testbed uses $2\times$ Intel Xeon 6330 CPUs with 512\,GB RAM and  NVIDIA A100 40\,GB GPU; in some experiments we replace the GPU with NVIDIA H200 or A40. 
The software stack includes Debian 12 with Linux kernel 6.1.0-17, Python 3.13, PyTorch 2.8, CUDA Toolkit 12.9, and NVIDIA driver version 12.3.

\subsection{Hyperparameter}
Table~\ref{tab:hyperparams} summarizes the key parameters used in the \name framework during end-to-end experiments.

\begin{table}[ht]
\centering
\caption{Detailed Hyperparameter Settings.}
\label{tab:hyperparams}
\small
\begin{tabular}{lll}
\toprule
\textbf{Parameter} & \textbf{Value} & \textbf{Description} \\
\midrule
Batch Size ($N$) & $256$ & Range of concurrent requests tested \\
Max Group Capacity ($\mathcal{C}$) & 8192 & Threshold for kernel packing efficiency \\
Regrouping Threshold & $\mathcal{C}/2$ & Accumulated drift condition to trigger re-balancing \\
Max New Tokens & 32 & Fixed for deterministic latency profiling \\
Paged Size & 128 \& 256 & Number of tokens per physical KV block \\
\bottomrule
\end{tabular}
\end{table}

\section{Solver Overhead}

We measure the grouping solver time and compare against an optimal formulation solved by Z3 theorem prover~\cite{Z3}. Our lightweight heuristic solver is orders of magnitude faster than the Z3-based optimum, making its overhead negligible relative to decoding latency.

It's worth noting that in the TTFT part, the experiment shows that Prepack can be $3\times$ worse when compared to \name.
A large part of the reason Prepack~\cite{Prepacking} performs poorly is its inefficient ILP solver. Through the heuristic solver in \name, we achieve both high performance and low overhead.

\begin{figure}[htbp]
	\centering
    \includegraphics[width=.4\linewidth]{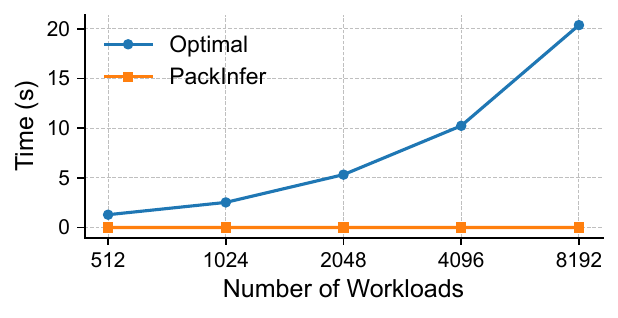}
	\caption{Solver time comparison}
	\label{fig:solver-time}
\end{figure}





\end{document}